\documentclass[prl,twocolumn,showpacs,nobibnotes,floatfix,superscriptaddress]{revtex4}

\usepackage{graphicx,eucal}

\begin{document}


\title{Quantum Dot Cluster State Computing with Encoded Qubits}
\author{Yaakov S. Weinstein}
\thanks{Present address: Quantum Information Science Group, MITRE, Eatontown, NJ, 07724\\ To whom correspondence should be addressed}
\email{weinstein@mitre.org}
\author{C. Stephen Hellberg}
\email{hellberg@dave.nrl.navy.mil}
\affiliation{Center for Computational Materials Science, Naval Research Laboratory, Washington, DC 20375}
\author{Jeremy Levy}
\email{jlevy@pitt.edu}
\affiliation{Dept. of Physics and Astronomy, University of Pittsburgh, Pittsburgh, PA, 15260 \bigskip}

\begin{abstract}
A class of architectures is advanced for cluster state quantum computation 
using quantum dots. These architectures include using single and multiple dots 
as logical qubits. Special attention is given to the supercoherent qubits 
introduced by Bacon, Brown, and Whaley [Phys. Rev. Lett. {\bf 87}, 247902 
(2001)] for which we discuss the effects of various errors, and present 
means of error protection.
\end{abstract}

\pacs{03.67.Lx, 
      03.67.Pp, 
      75.10.Jm} 
   
\maketitle

Cluster state or one-way quantum computation \cite{BR1} is a measurement based 
scheme for universal quantum computation formally equivalent to the traditional
circuit method \cite{BR2}. The basis of the scheme is the cluster state 
\cite{BR3}, a highly entangled state on which measurements alone can perform 
universal quantum computation. To create the cluster state one rotates all 
qubits in a lattice of at least two dimensions into the state 
$|+\rangle = 1/\sqrt{2}(|0\rangle+|1\rangle)$, followed by an Ising 
interaction $\exp(-i\frac{\pi}{4}\sigma_z^j\sigma^k_z)$ between all nearest 
neighbor qubits $j,k$ and single qubit $\sigma_z^j$ rotations.  

Due to the centrality of the Ising interaction in cluster state creation, 
proposed implementations of cluster state quantum computation 
\cite{N1,Browne,Zhang,Zeil,Natoms} have not included spin systems which 
evolve via the Heisenberg exchange interaction. However, these 
systems do allow for cluster state computation. In this work we discuss 
several cluster state computation models for systems with Heisenberg 
interaction. We outline the hardware requirements and possible error operators 
in each model. Encoding logical qubits (LQs) into several Heisenberg coupled 
physical qubits allows for a natural Ising-like inter-LQ evolution,
reduces hardware requirements, and allows for protection against various forms 
of errors and decoherence. While these models are applicable to any Heisenberg 
coupled systems we refer specifically to quantum dot implementations.

After creation of the cluster state, the desired algorithm is mapped onto the 
cluster state lattice by means outlined in \cite{BR1,BR2} and extra qubits are 
removed by measuring them along $z$. Computation is then performed by 
measurement of the remaining qubits along bases in the $x-y$-plane. For 
certain gates, such as arbitrary single qubit rotations, the outcome of a 
given measurement is necessary to determine in which basis future measurements 
should be performed \cite{BR1}. In such cases the measurements must be 
performed in a specified order.

For some systems, measurements along the $z$-axis \cite{Sasha} may be easier
than those in the $x-y$-plane. In this case one can rotate the qubit 
to be measured from the $x-y$-plane along $z$ and perform the measurement. 
For example, a process equivalent to measurement along an axis an angle $\phi$ 
away from positive $\hat{x}$, is a $\pi/2$-rotation about the axis 
$\phi-90^{\circ}$ and a measurement along $z$. Alternatively, all of the 
qubits can be rotated as long as they are rotated back after the measurement. 
We will make use of the commutivity between rotations and measurement in the 
models below.

\begin{figure}
\includegraphics[height=3.45cm]{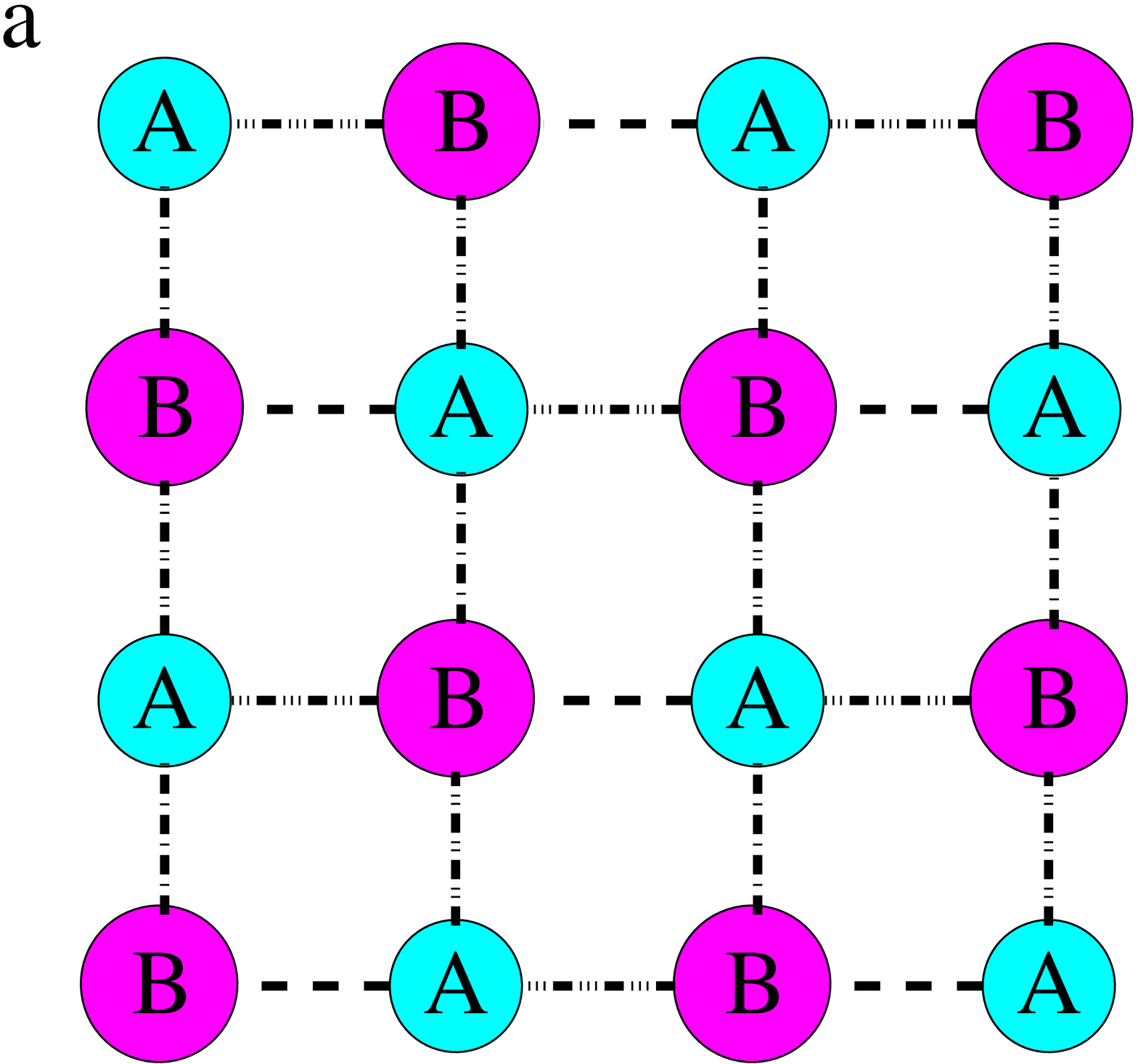} \hspace{.5cm}
\includegraphics[height=3.45cm]{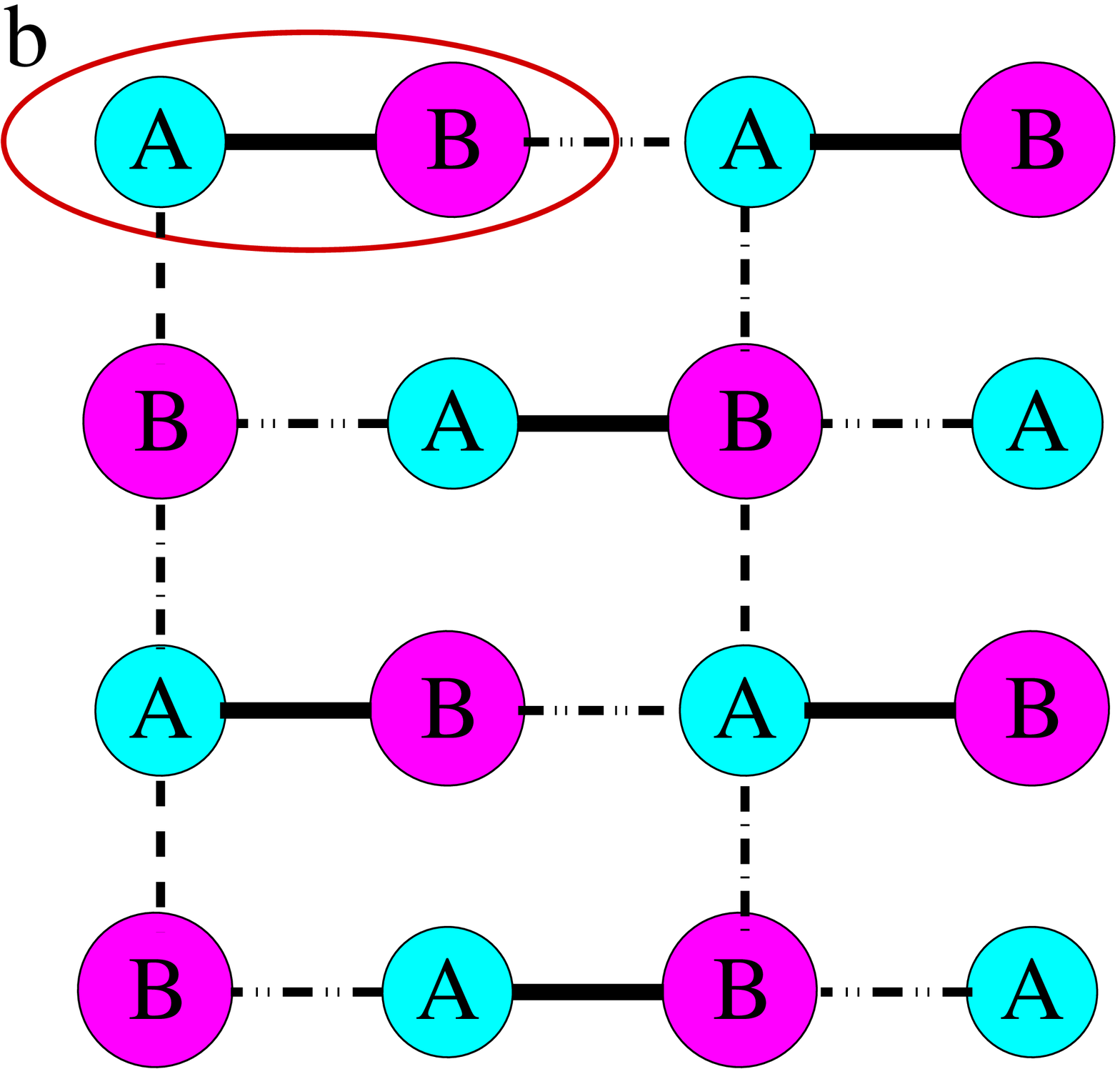}
\caption{\label{S1}
(Color online) Two species quantum dot cluster architectures. a) A two specie
cluster where each physical qubit, or dot, is a logical qubit. The cluster 
state for this architecture can be built in four coupling steps each requiring 
the application of $U_1$, $\pi)_z$ rotation on species $A$ or $B$, and again 
$U_1$. Four steps are necessary in order that a qubit will not interact with 
more than one neighboring qubit at each step. The connecting lines show which 
dots are coupled at each step: dashed, dash-dot, dash-dot-dot, and 
dash-dot-dot-dot, the order of the steps being unimportant. b) Each logical 
qubit is composed of two physical qubits, one from each species $A$ and $B$ 
(joined by solid lines). To build a cluster state one uses the intra-LQ 
coupling to create the $|+\rangle$ state on each qubit, and three coupling 
steps (dashed, dash-dot and dash-dot-dot lines) each consisting of inter-LQ 
coupling and single LQ $z$-rotations.}
\end{figure}

The first architecture we present for quantum dot cluster state computation is 
a two dimensional array of quantum dots, or physical qubits (PQs), each acting 
as a LQ. After initialization, the dots can be rotated into the state 
$|+\rangle$ using a global magnetic field about $\hat{y}$. For a pair of dots, 
an Ising interaction can be performed by evolution of the Heisenberg coupling, 
$U_1 = \exp(-i\frac{\pi}{8}\bf{\sigma}^1\cdot\bf{\sigma}^2)$, applying a 
$\pi$ rotation to one of the dots about $\hat{z}$, and again applying $U_1$. 
For an array of dots an Ising interaction between all nearest neighbors can
be implemented by turning on all nearest neighbor Heisenberg couplings and 
applying $\pi)_z$ rotations to every other dot. This requires individual 
addressability of the dots, via $g$-factor engineering or local magnetic 
fields, or alternative placement of two species of quantum dots ($A$ and $B$) 
in which all dots of a species are addressed equivalently and without 
effecting dots of the other species. We note that turning on all 
nearest-neighbor couplings at once introduces many-body terms into the 
Hamiltonian \cite{Mizel}. Instead, the cluster state can be built in four 
steps as in Fig.~\ref{S1}a. The construction is such that no qubit undergoes 
more then one coupling at each step. After the coupling terms single dot (or 
species of dot) rotations are needed to complete the cluster state.

Readout of the individual quantum dots can be accomplished via any of the 
suggested single-dot measurement schemes\cite{Burkard}. Certain of 
these schemes may be limited to measurement along the $z$-axis \cite{Sasha} 
in which case all of the qubits can be rotated (as explained above) using a 
global magnetic field.

Let us summarize the necessary components for this architecture. The hardware 
requirements are a two-dimensional lattice of individually addressable dots or 
alternative placement of two species of dots in which each species is 
addressed equivalently without affecting the other species. In addition, it 
is necessary to control the interactions between dots and perform single 
quantum dot measurements.

We can remove the necessity of performing logical operations with magnetic 
fields by encoding LQs into two physical qubits, one of each specie of dot
(where the species have different $g$-factors). Again the dots are arranged 
in a two dimensional lattice and placed in a magnetic field \cite{Levy}. The 
logical qubit states are:
\begin{eqnarray}
|0_L\rangle &=& |0_A1_B\rangle \nonumber\\
|1_L\rangle &=& |1_A0_B\rangle.
\end{eqnarray}
This encoding allows for the performance of single LQ  $x$-rotations by 
turning on the Heisenberg coupling between the physical qubits within the LQ 
(the intra-LQ coupling) and $z$-rotations by simply waiting and allowing the 
Zeeman term to evolve. Rotations about $\hat{z}$ require a time 
$\sim 1/B_z\Delta g$, where $B_z$ is the strength of the magnetic field and 
$\Delta g = g_A - g_B$ is the difference in species $g$-factor (an alternate 
method to perform $z$-rotations is given in \cite{Benj}). Inter-LQ interactions
are done via the Heisenberg coupling connecting two LQs. The latter operation 
takes part of the system state outside the LQ subspace. However, a logical 
$\pi)_z$ rotation on one of the LQs refocuses the unwanted part of the 
evolution leaving a logical Ising coupling, $\sigma^j_{z_L}\sigma^k_{z_L}$. In 
addition, this encoding is a decoherence free subspace (DFS) with respect to 
collective dephasing \cite{FV}.

With the above encoding a cluster state can be created as follows. Rotate all 
LQs into the $|+\rangle$ state via intra-LQ couplings. An Ising coupling 
between LQs is performed as described above. The coupling is applied in three
steps so as not to introduce multiple couplings on one physical qubit. Each 
step requires logical $\pi)_z$ rotations on alternate LQs only. This can be 
done by via the quicker logical $\pi)_x$ rotations on the remaining LQs. 
Finally, single LQ rotations are done using the Zeeman interaction. The 
required lattice for this architecture and the scheme for cluster state 
construction is shown in Fig.~\ref{S1}b. Measurement of the LQs can be done 
by measuring the individual quantum dots as above or directly \cite{Levy} via 
singlet-triplet measurements \cite{Kane}.

To summarize, the requirements for cluster state computation using LQs 
encoded in two physical qubits are two species of quantum dots placed 
alternately in a two-dimensional lattice. The two species $g$-factors 
should be different enough such that LQ $z$ rotations can be done
within a reasonable time. In addition, it is necessary to control the 
interactions between dots and perform either single dot or singlet-triplet 
measurements.

The previous cluster state constructions all require the utilization of 
more than one species of physical qubit. By incorporating more advanced 
logical qubits we can remove this necessity, perform computations using just 
the Heisenberg interaction and singlet-triplet measurement, while making the 
LQs more robust against a variety of possible errors. In addition, 
cluster state construction becomes easier since the Ising interaction is the 
natural coupling between the logical qubits. An example 
of this type of DFS is the three physical-qubit encoding discussed in 
Ref. \cite{YSW} which provides DFS protection against collective errors. The 
inter-LQ Hamiltonian for this three-qubit encoding is 
$\sigma_{z_L}^j\sigma_{z_L}^k$ plus single LQ $z$-rotations which can then be 
modified by intra-LQ couplings.

A more robust encoding is the supercoherent qubit (SQ) of Ref. \cite{superco}. 
SQs minimize decoherence by forcing all reasonable interactions with the 
environment to overcome an energy gap which exists between the logical qubit 
subspace and other states of the system. Environment induced collective errors 
or single physical qubit local errors must overcome this energy gap. 

A SQ in its idle state consists of four physical qubits with equal non-zero 
Heisenberg coupling (equal to 1 for simplicity) between all physical qubit
pairs. An orthogonal basis for the doubly degenerate ground state comprises of 
singlet or triplet states between physical qubits 1,2 and physical qubits 3,4:
\begin{eqnarray}
|0_L\rangle &=& \frac{1}{2}\left(|\uparrow\downarrow\rangle-|\downarrow\uparrow\rangle\right)_{1,2}\otimes\left(-|\uparrow\downarrow\rangle+|\downarrow\uparrow\rangle\right)_{3,4}\nonumber \\
|1_L\rangle &=& \frac{1}{\sqrt{3}}\left(|\uparrow\uparrow\downarrow\downarrow\rangle+|\downarrow\downarrow\uparrow\uparrow\rangle\right) \nonumber\\
&-& \frac{1}{2\sqrt{3}}\left(|\uparrow\downarrow\rangle+|\downarrow\uparrow\rangle\right)_{1,2}\otimes\left(|\uparrow\downarrow\rangle+|\downarrow\uparrow\rangle\right)_{3,4}.
\end{eqnarray}
The state $|0_L\rangle$ incorporates the singlet state between spins 1 and 2 
and spins 3 and 4. The state $|1_L\rangle$ incorporates triplet states between 
the same spins. We will refer to the pairs of spins 1-2 and 3-4 as `singlet' 
pairs since they are in a singlet state when the LQ is in the logical state 
$|0_L\rangle$. Measurement of the SQ can be accomplished via 
singlet-triplet measurement schemes \cite{Kane}. Initialization of the SQ is 
done by raising or lowering one of the couplings, thereby breaking the ground
state degeneracy, and waiting for the system to decay \cite{Div}. 

Single SQ rotations are performed by changing the Heisenberg coupling 
strength between appropriate pairs of qubits. Changing a coupling between a 
qubit and its `singlet' pair, 1 and 2 for example, performs a logical $z$ 
rotation. Changing a coupling between a qubit and one that is not its pair, 
for example 2 and 3, performs a rotation about the axis in the $x-z$ plane 
$120^{\circ}$ from the $z$ axis \cite{Div,superco,YSW}. Combinations of these 
operations are sufficient to perform any $SU(2)$ rotation.

To perform inter-SQ interactions Ref.~\cite{superco} considers 
eight physical qubits with equal Heisenberg coupling between all pairs. In 
this way interactions between the four-qubit SQ can be implemented and 
universal quantum computation can proceed without leaving the LQ subspace. 
The problem with this coupling scheme is its realization. As already 
noted in \cite{superco}, building two coupled SQs would be a daunting task 
and scaling up will only add complications.

A simplified scheme for implementing inter-SQ couplings is introduced in 
\cite{YSWEx4}. The basis of this scheme is that if the inter-SQ interactions 
are performed adiabatically \cite{YSW} the system returns to the logical 
subspace when the couplings between SQs are turned off. This condition is not 
difficult to satisfy. Adiabatic evolution is a requirement for {\em all} 
approaches using spins in quantum dots \cite{SLM}. A coupling between one 
physical qubit in each logical qubit does not induce any evolution of the SQ, 
since SQs are robust against local operations. Instead, coupling between pairs 
of physical qubits are necessary. The pairs which give Ising coupling between 
SQs are those that are in a singlet (triplet) in the state $|0_L\rangle$ 
($|1_L\rangle$), such as 1 and 2. When pairs of such qubits in two SQs are 
coupled the resulting SQ coupling is diagonal consisting of 
$\sigma_{z_L}^j\sigma_{z_L}^k$ and equal single-SQ logical $\sigma_{z_L}$ 
rotations. The two-SQ interaction plus equal logical $z$-rotations on each of 
the two SQs can implement an Ising interaction which, combined with arbitrary 
single SQ rotations, allow for a universal set of gates \cite{Bar}.

\begin{figure}
\includegraphics[height=3.2cm]{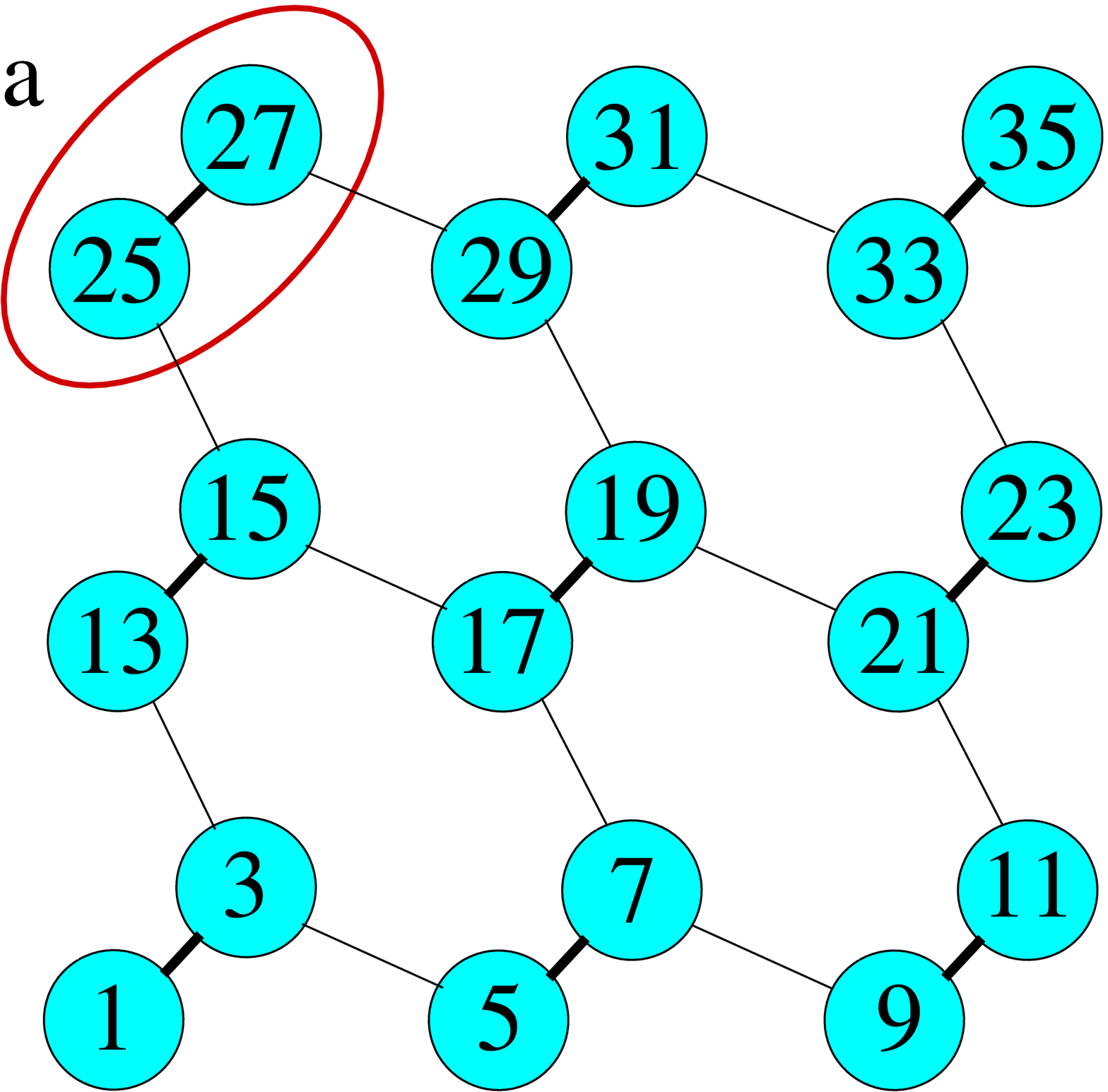} \hspace{1cm}
\includegraphics[height=3.2cm]{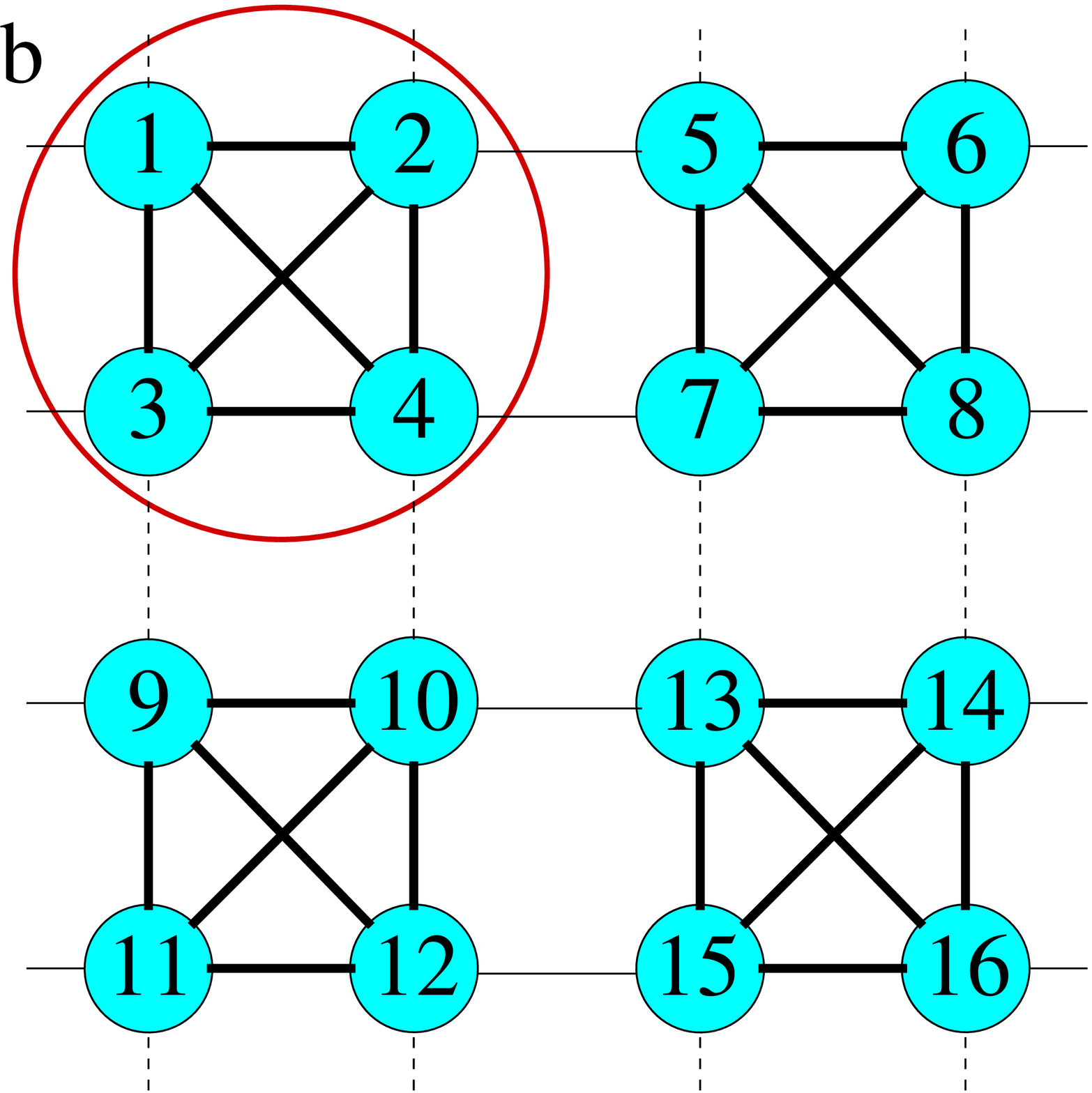}
\caption{\label{Dots}
(Color online) Proposed layouts for a two-dimensional cluster state made of 
supercoherent qubits (SQs). Thick lines represent the always-on exchange 
interaction which exists between all pairs within a single SQ. These are 
modified to perform single LQ rotations. Thin lines represent exchange 
interactions between SQs which are turned on to perform coupling between the 
SQs. Two couplings are necessary. a) Two layer architecture with only top 
layer shown. Top layer consists of two of the four physical qubits (odd 
numbered) per SQ. The two even numbered physical qubits are placed directly 
beneath the odd numbered ones. Couplings exist between all pairs within a SQ. 
b) Four SQs on a two-dimensional lattice requiring two steps to cluster state 
creation. First, inter-SQ couplings are turned on between nearest horizontal 
SQs (solid lines). This creates a cluster state across each row of the lattice.
Then a SWAP is applied between physical qubits at the ends of a diagonal within
each SQ (for example 1 and 4) and the coupling is turned on between horizontal 
nearest neighbors (dashed lines). Another SWAP and single SQ $z$-rotations 
(done by intra-SQ coupling) complete the required inter-SQ evolution.
}
\end{figure}

To create the cluster state we need to arrange the SQs in two-dimensions.
This is complicated by the fact that only couplings between certain pairs of 
physical qubits lead to a diagonal coupling for the LQs. We suggest two designs
which allow the appropriate qubits to be coupled. The first requires arranging 
dots in two layers. The top layer of such an architecture is shown in Fig. 
\ref{Dots}a and the second layer is placed directly underneath. Each SQ 
consists of two physical qubits on each layer with identical couplings
between all physical qubits.

An alternative is to arrange the SQs in a two-dimensional lattice as in Fig. 
\ref{Dots}b. The Ising interaction between all SQs is done as follows. 
Assume that the singlet pairs are horizontal (1-2, 3-4). Then the vertical 
couplings between the SQs (3-9, 4-10) can be turned on to implement the Ising 
interaction (plus single SQ $z$-rotations). We then apply a SWAP operation 
between dots on the diagonal of each SQ (1-4, 5-8). The SWAP is implemented by 
increasing the Heisenberg interaction between those qubits for a time 
$t = \pi\Delta J$ where $\Delta J$ is the amount the coupling is changed from 
its original value. The singlets are now vertical (1-3, 2-4) and turning on 
the vertical couplings (2-5, 4-7) implements an Ising interaction between the 
horizontally arranged SQs. Another SWAP between physical qubits along the 
diagonal of the SQs completes the interactions.

To summarize, encoding logical qubits into three and four physical qubits 
allows for cluster state computation with only one species of quantum dot. 
These logical qubits are robust against a variety of decoherence mechanisms
and allow for universal cluster state computation without using magnetic 
fields to manipulate qubits. The challenge is the arrangement of LQs in two
dimensions and the ability to change the coupling strength between physical 
qubits. 

The three and four qubit encodings are powerful means of 
protecting quantum information against a variety of errors in both circuit 
model and cluster state quantum computation. We now present a 
basic examination of errors that may occur when cluster state computation 
is performed with these architectures and possible methods of correction.

The three- and four-qubit encodings require equal couplings between multiple 
qubits. This is exceedingly difficult because the coupling strength depends 
exponentially on the spacing between the electrons \cite{Fries}. For the 
four-qubit encoding the challenge is greater since one cannot put the 
dots at equal distances from each other in just two dimensions. Methods of 
dealing with this latter issue are given in \cite{YSWEx4}. The effect 
of having an unequal coupling in one of the LQs is a constant, linear, 
single-LQ rotation similar to the chemical shift term in liquid-state nuclear 
magnetic resonance (NMR) \cite{PhysD}. A possible error correcting technique 
is open-loop control or bang-bang operations \cite{BB}. As in NMR 
decoupling schemes, logical $\pi$-pulses can be applied to refocus unwanted 
evolution. A hardware approach is to digitize the interaction between qubits 
by enforcing minimum and maximum coupling strengths and a smooth transition 
between them. One method of achieving a maximum coupling is by moving electrons
transversely in channels \cite{Fries}. Another method is to use two ancilla 
quantum dots in an arrangement known as a quantum gate circuit (QGC)
\cite{H}. QGCs can provide a truly digital coupling, insuring zero coupling 
and a maximum value, thereby easing the process of making the couplings equal. 
Either of these hardware based designs allows for more control of the 
couplings between physical qubits easing the task of setting the different
couplings to equal values, but require the addition of more dots per LQ.

Inter-LQ couplings provide another possible source of errors. For both the 
three and four physical qubit encodings there is the problem of insuring the 
coupling is exactly zero when required. Achieving zero coupling can be done to 
some accuracy through electro-static gates and improved by bang-bang decoupling
or via QGCs \cite{H} which guarantee zero coupling. 

Inter-SQ evolution requires two adiabatic couplings between pairs of 
physical qubits. Errors that may occur include the ability to turn on the 
coupling simultaneously and the possibility that the coupling strengths 
are unequal. As SQ are immune to single physical qubit errors, having one 
of the couplings on will not effect the state of the SQs. Thus, the task
of turning the couplings on at the same time is only an issue during 
the ramp up when the couplings may be unequal. For unequal coupling strengths 
the one-SQ $\sigma_z$ terms and the two-SQ $\sigma_z^j\sigma_z^k$ terms depend
linearly on the difference in coupling strengths. No other terms are introduced
due to the inequality. This error can be corrected with LQ  $\pi$-pulses or 
QGCs can be set up to help insure equal coupling.

In conclusion, we have suggested several ways of achieving cluster state 
quantum computation using quantum dots. The architectures range from having
one to four physical qubits per logical qubit. We have also analyzed possible 
errors for the three and four qubit encodings and suggested ways of fixing 
them. Which architecture is most practical depends on the particular physical  
implementation.

The authors acknowledge support from the DARPA QuIST (MIPR 02 N699-00)
program. Y.S.W. acknowledges support of the National Research Council
through the Naval Research Laboratory. Computations were performed at the
ASC DoD Major Shared Resource Center.

\end{document}